\documentclass[aps,a4paper,10pt,twocolumn,nofootinbib]{revtex4} % eqsecnum
\usepackage[T1]{fontenc}
\usepackage{mathptmx}
\usepackage{datetime}
\usepackage{amsmath}
\usepackage{amsfonts}
\usepackage{amsfonts}
\usepackage{mathrsfs}
\usepackage[mathscr]{euscript}
\usepackage[dvips]{graphicx}
\usepackage{fancyhdr}
\usepackage{colordvi}
\usepackage[hypertex=true]{hyperref}
\usepackage{epsfig}
\usepackage{color}
\usepackage{bm}

\def\overstrike#1#2{{\setbox0\hbox{$#2$}\hbox to \wd0{\hss
    $#1$\hss}\kern-\wd0\box0}}

\renewcommand{\Vec}{\bm}

        \DeclareMathOperator{\grad}{\nabla}

\newcommand{\XDOI}[1]{\href{http://dx.doi.org/#1}{doi:#1}}
\newcommand{\XARXIV}[1]{\href{http://arxiv.org/abs/#1}{arXiv:#1}}

\newdateformat{yymmdddate}{\THEYEAR/\twodigit{\THEMONTH}/\twodigit{\THEDAY}}

\begin{document}
\title{Deriving the time-dependent Schr\"{o}dinger $m$-
 and $p$-equations from the Klein-Gordon equation.}
\author{Paul Kinsler}
\email{Dr.Paul.Kinsler@physics.org}
\affiliation{
  %Blackett Laboratory, Imperial College,
  Department of Physics, 
  Imperial College London,
  Prince Consort Road,
  London SW7 2AZ, 
  United Kingdom.
}

\lhead{\includegraphics[height=5mm,angle=0]{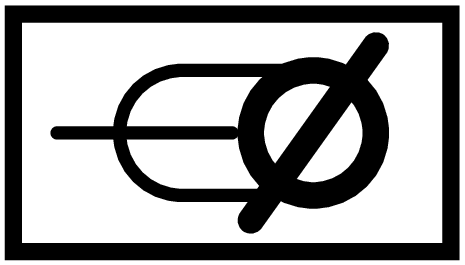}~~KG2SCHRO}
\chead{Deriving Mass Schr\"{o}dinger}
\rhead{
\href{mailto:Dr.Paul.Kinsler@physics.org}{Dr.Paul.Kinsler@physics.org}\\
\href{http://www.kinsler.org/physics/}{http://www.kinsler.org/physics/}
}

\begin{abstract}

I present an alternative and rather direct way to derive
 the well known Schr\"odinger equation for a quantum wavefunction, 
 by starting with the Klein Gordon equation 
 and applying a directional factorization scheme.
And since if you have a directionally factorizing hammer, 
 everything looks like a factorizable nail, 
 I also derive 
 an alternative wavefunction propagation equation
 in the momentum-dominated limit.
This new Schr\"odinger $p$-equation therefore provides
 a potentially useful complement
 to the traditional Schr\"odinger $m$-equation's 
 mass-dominated limit.

\end{abstract}

%\pacs{}

\date{\today}
\maketitle
\thispagestyle{fancy}

%
% =======================================================================
\section{Introduction}\label{S-intro}

There have been many and varied (re)derivations
 of the Schr\"odinger equation \cite{Briggs-R-2001fp}, 
 based on a variety of principles -- 
 e.g. Feynman path integrals \cite{Derbes-1996ajp}, 
 stochastics (e.g. \cite{Nelson-1966pr,Davies-1989jpa}),
 utilizing axioms \cite{Sanayei-2013arXiv},
 or by applying various ad hoc approximations 
 to variants of the Klein-Gordon equation (e.g. \cite{Ward-V-2006arXiv}).
Here I present another method, 
 inspired by the success of directionally-based
 factorizations of optical wave equations \cite{Kinsler-2010pra-fchhg}, 
 which allow us to proceed whilst 
 making only the bare minimum of approximations.
Of course, 
 one might say that the approximation used here 
 to (re)derive the Schr\"odinger equation is the usual one, 
 and so little has been achieved.
However, 
 as in its applications in optics
 \cite{Kinsler-2010pra-fchhg,Kinsler-2012arxiv-fbrad}
 and acoustics \cite{Kinsler-2012arXiv-fbacou}, 
 the gains are threefold:

\begin{enumerate}

\item
% first
 The Klein-Gordon equation
 is recast \emph{without approximation} into a new form 
 designed to isolate the part intended to be approximated away, 
 making the nature of the approximation much clearer. 

\item
% second, 
 That new form enables us to compare in all details
 the exact and approximate versions side by side, 
 either mathematically or computationally.

\item
%Thirdly, 
 The method encourages us to explore alternate approximations -- 
 here, 
 a momentum-dominated limit complementary
 to the traditional mass-dominated one used to obtain the 
 ordinary Schr\"odinger equation.

\end{enumerate}

Klein \& Gordon 
 started with the relativistic equation for the energy
 of a massive particle, 
~
\begin{align}
 E^2
&=
  m^2 c^4 + p^2 c^2
,
\label{eqn-massiveenergy}
\end{align}
 and, 
 by replacing $E$ and $p$ with operators
 using the correspondence principle \cite{Bohr-TheCP}.
From a mathematical perspective, 
 the correspondence principle
 is just the process of switching
 between one domain and its Fourier transformed counterpart.
Here, 
 the correspondence is
~
\begin{align}
  E       & \leftrightarrow \imath \hbar \partial_t
,
\\
  \Vec{p} & \leftrightarrow - \imath \hbar \grad
,
\label{eqn-correspondence}
\end{align}
 which allows us to directly convert eqn. \eqref{eqn-massiveenergy}
 into the Klein-Gordon (KG) equation for 
 the wavefunction of a single massive particle,
 i.e. 
~
\begin{align}
  \left[
    \hbar^2
    \partial_t^2
   +
    m^2 c^4
   -
    \hbar^2 c^2 \grad^2
  \right]
  \Phi(\Vec{r},t)
&=
  0
,
\label{eqn-KG-1}
\\
  \textrm{or} \qquad \qquad
  \left[
    \frac{1}{c^2}
    \partial_t^2
   +
    \frac{m^2 c^2}{\hbar^2}
   -
    \grad^2
  \right]
  \Phi(\Vec{r},t)
&=
  0
.
\label{eqn-KG-2}
\end{align}
This Klein-Gordon \emph{second order} wave equation can, 
 if desired, 
 be factorized using spinors 
 to give the first order Dirac equation.
However, 
 this does not allow for anything that might alter the wavefunction
 behaviour away from that in a simple vacuum, 
 so to address this lack I consider modifications
 inspired by both the both the Salpeter Hamiltonian
 and a gravitiational potential.

%
% -----------------------------------------------------------------------
%\subsection{The Salpeter Hamiltonian}\label{SS-potSalt}

\noindent
\textit{The Salpeter Hamiltonian:~}
It is useful -- 
 especially when deriving the Schr\"odinger equations --
 to be able to include the effect of a static potential
 within which the particle is moving.
We might therefore start with the Salpeter Hamiltonian \cite{Salpeter-B-1951pr}
~
\begin{align}
 H \Phi &= \left[ \sqrt{m^2 c^4+p^2c^2}+V(r) \right]  \Phi
,
\end{align}
 where the Hamiltonian can be applied twice 
 to the wavefunction $\Phi$.
Then,  
 by identifying $H$ with the energy $E$, 
 we get the squared form
~
\begin{align}
 \left( E - V \right)^2  \Phi(\Vec{r},t)
&=
  \left[ m^2 c^4 + p^2 c^2 \right] \Phi(\Vec{r},t)
,
\label{eqn-salpeter2}
\end{align}
 which matches up to the Klein-Gordon starting point
 under the condition that $V=0$.
As would be expected, 
 the same as the Klein-Gordon equation in a Coulomb potential
 if $V=-e^2/r$.
In the following, 
 I will call the potential $V$ the ``Salpeter potential''
 to specify its conceptual origin.

%
% -----------------------------------------------------------------------
%\subsection{Gravitational potential}\label{SS-potGRN}

\noindent
\textit{Gravitational potential:~}
Although it might seem unlikely that gravitational potentials 
 have sufficient variation in either space or time to produce
 effects that apply to quantum phenomena, 
 it is nevertheless interesting to see how gravity might appear in
 the Schr\"odinger equation.
In general relativity,
 the Newtonian limit for a gravitational potential $\Xi(\Vec{r},t)$ 
 gives an expression for $E^2$ which is
 \cite{Schutz-FCRelativity,Carroll-LNGenRel}
%}\footnote{See 
% arXiv:gr-qc/9712019, p.106 (pdf.113), eqns. (4.21), (4.22).}.
~
\begin{align}
  E^2
&=
  m^2 c^4
  \left[ 1 + 2 \Xi(\Vec{r},t) \right]
+
  p^2 c^2
.
\label{eqn-energy2-gravity}
\end{align}
In an operator form, 
 applied to some wavefunction $\Phi$, 
 this would then be
~
\begin{align}
  E^2 \Phi(\Vec{r},t)
&=
  \left\{
    m^2 c^4
    \left[ 1 + 2 \Xi(\Vec{r},t) \right]
   +
    p^2 c^2
  \right\}
  \Phi(\Vec{r},t)
.
\label{eqn-energy2-gravity2}
\end{align}

Since there is no elegant way to handle the \emph{time dependent} $\Xi$ term
 elegantly as part of the energy
 (i.e. on the LHS),
 it is best left as a perturbation 
 and treated in the same way as the momentum --
 both are small in the non-relativistic limits.
Note that when properly scaled, 
 we can also use $\Xi(\Vec{r},t)$ as a proxy for any other 
 space and time dependent potential that might 
 affect our system.

%
% -----------------------------------------------------------------------
%\subsection{Combined potentials}\label{SS-potBoth}

\noindent
\textit{Combined potentials:~}
So we only have to perform the following calculation \emph{once}, 
 I will combine the Salpeter energy expression eqn. \eqref{eqn-salpeter2} with 
 that allowing for a gravitational potential eqn. \eqref{eqn-energy2-gravity}.
For an operator-like form, 
 applied to a wavefunction $\Phi$, 
 we have
~
\begin{align}
  \left[
    E -  V(\Vec{r})
  \right]^2
  \Phi(\Vec{r},t)
&=
  \left\{
    m^2 c^4
    \left[ 1 + 2 \Xi(\Vec{r},t) \right]
   +
    p^2 c^2
  \right\}
  \Phi(\Vec{r},t)
\\
&=
  \left[
  m^2 c^4
+
  2 m^2 c^4 \Xi(\Vec{r},t)
+
  p^2 c^2
  \right]
  \Phi(\Vec{r},t)
.
\label{eqn-energy2-SalGrav}
\end{align}
Here the potential $V(\Vec{r})$ has no $t$ dependence, 
 because allowing that would complicate the transformation
 of the LHS into a time derivative.
However, 
 any time dependent part of a more general $V$ could 
 easily be merged into $\Xi(\Vec{r},t)$.
In a Klein-Gordon wave equation form, 
 this is
~
\begin{align}
  \left\{
    \frac{1}{c^2}
    \partial_t^2
   +
    \frac{m^2 c^2}{\hbar^2}
    \left[
     1 + 2 \Xi(\Vec{r},t)
    \right]
   -
    \grad^2
  \right\}
  \Phi(\Vec{r},t)
&=
  0
\label{eqn-Helmholtz-SalGrav}
.
\end{align}
In frequency space
 ($\omega, \Vec{r}$-space), 
 this becomes
~
\begin{align}
  \left\{
   -
    \frac{\omega^2}{c^2}
   +
    \frac{m^2 c^2}{\hbar^2}
    \left[
     1 + 2 \breve{\Xi}(\Vec{r},\omega)
    \right]
   -
    \grad^2
  \right\}
  \Phi(\Vec{r},\omega)
&=
  0
\label{eqn-Helmholtz-SalGrav-rw}
,
\end{align}
 where the breve (here $\breve{\Xi}$)
 tells us to convolve $\Xi$ with $\Phi$ over $\omega$.
Alternatively, 
 in wavevector space
 ($t, \Vec{k}$-space), 
 this becomes
~
\begin{align}
  \left\{
    \frac{1}{c^2}
    \partial_t^2
   +
    \frac{m^2 c^2}{\hbar^2}
    \left[
     1 + 2 \hat{\Xi}(\Vec{k},\omega)
    \right]
   +
    k^2
  \right\}
  \Phi(\Vec{k},\omega)
&=
  0
\label{eqn-Helmholtz-SalGrav-kt}
,
\end{align}
 where the hat (here $\hat{\Xi}$)
 tells us to convolve $\Xi$ with $\Phi$ 
 over $\Vec{k}$.
Both types of convolution play no interesting role
 in the following calculations, 
 and are merely an intermediate stage which disppears
 when the equations being used are converted
 back into their primary $t, \Vec{r}$ domain.

%
% -----------------------------------------------------------------------
%\subsection{Method}\label{SS-method}

\noindent
\textit{Method:~}
In what follows, 
 I use eqn. \eqref{eqn-Helmholtz-SalGrav}
 which contains two different types of potential,
 to derive approximate equations
 which have only first order derivatives in the propagation variable; 
 i.e. $t$ for the usual temporally propagated Schr\"odinger equation.
To complement the Schr\"odinger equation derivation, 
 I also derive
 a spatially-propagated version, 
 which is applicable in a different limit.
For a more systematic look at the differences between 
 temporal propagation
 and spatial propagation, 
 the reader is referred to Ref. \cite{Kinsler-2012arXiv-fbacou}.
Further, 
 although here we factorize in Cartesian coordinates, 
 this is not the only possible choice \cite{Kinsler-2012arxiv-fbrad}.
Finally, 
 note that my original source for the factorization method used
 was by Ferrando \textit{et al.} \cite{Ferrando-ZCBM-2005pre}.

%
% =======================================================================
\section{Mass dominant: the Schr\"odinger equation}\label{S-tshrodinger}

We can see from the correspondence principle described above
 that the energy $E$ is related to evolution in time $t$, 
 while also noting that in non-relativistic scenarios 
 the bulk of a massive particle's energy
 is frozen in its rest mass.
Thus to reduce the second-order-in-time KG equations
 down to the first-order-in-time Schr\"odinger equation
 we need to manipulate the starting equations 
 while focussing on the energy $E$, 
 and the rest mass $m$.

\begin{figure}
 \includegraphics[width=0.80\columnwidth,angle=0]{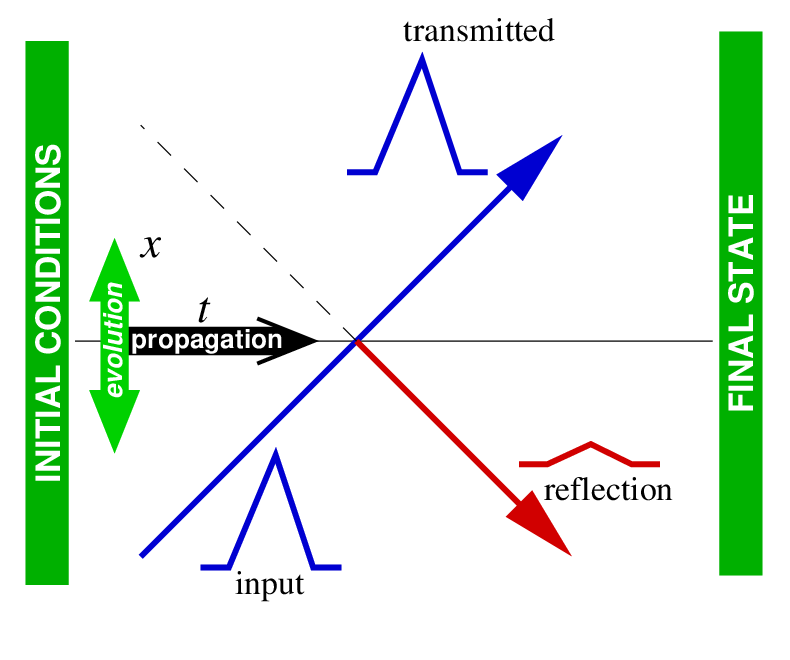}
%{propagator}
\caption{
For temporal propagation (right), 
 initial conditions cover all space
 at an initial time $t_i$; 
 the final state at $t_f$ also covers all space.
The effect of a reflective interface
 is also indicated, 
 since it makes the important distinction between 
 \emph{propagation} and \emph{evolution} clearer.
}
\label{F-diagram-tpropagator}
\end{figure}

To proceed I will follow the directional factorization method 
 recently popularized in optics
 \cite{Kinsler-2010pra-fchhg},
 albeit with an alternate physical focus on temporal propagation
 (see e.g. \cite{Kinsler-2012arXiv-fbacou}).
This is the most physically motivated factorization,
 and we  
 decompose the system behaviour (waves) into directional components
 that then evolve either forward or backward in space, 
 as shown in Fig. \ref{F-diagram-tpropagator}.
To analyse temporal propagation, 
 we need a useful reference paramater to characterise it,
 and it should preferably be one that remains constant.
In this case, 
 a frequency domain analysis is called for:
 we might therefore use either an energy or a frequency $\omega$.
This means that the parts of the physics we wish to ascribe 
 to the role of ``reference propagation'' must be time independent.

Start by defining $E' = E - V(\Vec{r}) = \hbar \omega$
 to work in a scaled frequency ($\omega$) space,
 so that 
~
\begin{align}
 \hbar^2 \omega^2
  \Phi(\Vec{r},\omega)
&=
 \left[
   m^2 c^4 
  +
    2 m^2 c^4 \breve{\Xi}(\Vec{r},\omega)
   -
    \hbar^2 c^2 \grad^2
  \right]
  \Phi(\Vec{r},\omega)
\end{align}
%and
%~
\begin{align}
 \left[
  \hbar^2 \omega^2 - m^2 c^4 
  \right]
  \Phi
&=
  \left[
    2 m^2 c^4 \breve{\Xi}
   -
    \hbar^2 c^2 \grad^2
  \right]
  \Phi
\\
  \left( \hbar \omega - mc^2 \right)
  \left( \hbar \omega + mc^2 \right)
  \Phi
&=
  \left[
    2 m^2 c^4 \breve{\Xi}
   -
    \hbar^2 c^2 \grad^2
  \right]
  \Phi
\\
  \Phi
&=
  \frac{\breve{Q}}
       {\left( \hbar \omega - mc^2 \right) \left( \hbar \omega + mc^2 \right)}
  \Phi
\\
  \Phi
&= 
  \left[
    \frac{1/2mc^2}
         {\hbar \omega - mc^2}
   -
    \frac{1/2mc^2}
         {\hbar \omega + mc^2}
  \right]
  \hat{Q}
  \Phi
,
\label{eqn-tprop-mainderivH}
\\
\textrm{where} \qquad
  \breve{Q}(\Vec{r},\omega)
&= 
  2 m^2 c^4 \breve{\Xi}(\Vec{r},\omega) - \hbar^2 c^2 \grad^2
.
\end{align}

We can see from the term in square brackets % $[]$
 on the RHS of eqn. \eqref{eqn-tprop-mainderivH}
 that $\Phi$ evolves according to two complementary parts
 of differing sign.
The term proportional to $(\hbar\omega - mc)^{-1}$
 generates a forward-like evolution, 
 and that proportional to $(\hbar\omega + mc)^{-1}$
 generates a backward-like evolution
 \cite{Ferrando-ZCBM-2005pre}.
As a result we can likewise split the wavefunction
 into corresponding pieces, 
 with $\Phi \equiv \Phi_+ + \Phi_-$.
When we transform back into the time domain, 
 these will (must!) propagate forward in time $t$, 
 all the while holding information about the wavefunction
 as a function of $\Vec{r}$.
To avoid notational clutter, 
 we use this fact as an excuse to omit the time argument, 
 and only the $\Vec{r}$ argument of $\Phi_\pm$ will be given.

Further, 
 since the $\Phi_+$ forward evolving component is by definition
 propagating to later times, 
 its excitations therefore must (also) be understood to be 
 evolving forward in space ($\Vec{r} \rightarrow \infty$).
In contrast, 
 the $\Phi_-$ backward evolving component
 (also propagating to later times), 
 therefore has excitations
 evolving backward in space ($\Vec{r} \rightarrow -\infty$).

Continuing the separation of $\Phi_+$ and $\Phi_-$, 
 we see that 
~
\begin{align}
  \Phi_+(\Vec{r}) + \Phi_-(\Vec{r})
&=
  \left[
    \frac{1/2mc^2}
         {{\hbar \omega} - mc^2}
   -
    \frac{1/2mc^2}
         {{\hbar \omega} + mc^2}
  \right]
  \breve{Q}
  \Phi(\Vec{r})
\\
  \Phi_\pm(\Vec{r})
&=
  \pm
    \frac{\left(2 m^2 c^4 \breve{\Xi} - \hbar^2 c^2 \grad^2 \right)}
         {\left(2mc^2\right) ~~ \left({\hbar \omega} \mp mc^2\right)}
  \left[
    \Phi_+(\Vec{r}) + \Phi_-(\Vec{r})
  \right]
\\
  \Phi_\pm(\Vec{r})
&=
  \pm
    \frac{\left(m c^2 \breve{\Xi} - \frac{\hbar^2\grad^2}{2m}\right)}
         {{\hbar \omega} \mp mc^2}
  \left[
    \Phi_+(\Vec{r}) + \Phi_-(\Vec{r})
  \right]
,
\end{align}
 which enables us to write
~
\begin{align}
  \left(
    {\hbar \omega} \mp mc^2
  \right)
  \Phi_\pm(\Vec{r})
&=
  \pm
  \left(
     mc^2 \breve{\Xi}
    -
     \frac{\hbar^2\grad^2}{2m}
  \right)
  \left[
    \Phi_+(\Vec{r}) + \Phi_-(\Vec{r})
  \right]
\\
  \left(
    E
   -
    V
   \mp
    mc^2
  \right)
  \Phi_\pm(\Vec{r})
&=
  \pm
  \left(
     mc^2 \breve{\Xi}
    -
     \frac{\hbar^2\grad^2}{2m}
  \right)
  \left[
    \Phi_+(\Vec{r}) + \Phi_-(\Vec{r})
  \right]
,
\end{align}
 and finally
~
\begin{align}
  E  
  \Phi_\pm(\Vec{r})
&=
 \pm
  mc^2
  \Phi_\pm(\Vec{r})
 +
  V 
  \Phi_\pm(\Vec{r})
\nonumber
\\
&\qquad
 \pm
  \left(
     mc^2 \breve{\Xi}
    -
     \frac{\hbar^2\grad^2}{2m}
  \right)
  \left[
    \Phi_+(\Vec{r}) + \Phi_-(\Vec{r})
  \right]
.
\end{align}
If $\Phi_{-}$ is set to zero, 
 and only the $\Phi_{+}$ is considered, 
 we see that in terms of momentum $p = \hbar k$, 
 and for a time-independent $\Xi$, 
 this will have the %non-relativistic 
 dispersion relation
 $E = E(p) = mc^2 + V + mc^2 \Xi(k) + p^2/2m$.

By using the  correspondence principle \cite{Bohr-TheCP}
 to replace
~
\begin{align}
  E       & \leftrightarrow \imath \hbar \partial_t
,
\end{align}
 we see that back in $t,\Vec{r}$ space the
 convolution vanishes, 
 and we get a pair of coupled differential equations,
~
\begin{align}
  \imath
  \hbar
  \partial_t
  \Phi_\pm(\Vec{r})
&=
 \pm
  mc^2
  \Phi_\pm(\Vec{r})
 +
  V 
  \Phi_\pm(\Vec{r})
\nonumber
\\
&\qquad
 \pm
  \left(
     mc^2 \Xi
    -
     \frac{\hbar^2 \grad^2}
          {2m}
  \right)
  \left[
    \Phi_+(\Vec{r}) + \Phi_-(\Vec{r})
  \right]
.
\end{align}
This is a pair of first order wave equations coupled 
 only by the gravitational potential $\Xi(\Vec{r},t)$
 and the momentum squared term 
 (i.e. that $\propto \grad^2$); 
 the potential $V$ does not couple the two because 
 it was chosen to be time independent.
Those couplings,  
 along with the rest mass, 
 the wavefunction(s), 
 and their spatial derivatives,
 then tell us how $\Phi_\pm(\Vec{r})$ will change
 on propagating forward in time.

If both $\Xi$ and the momentum are small compared
 to the (dominant) mass term, 
 as is true in Newtonian and non-relativistic scenarios, 
 then any finite $\Phi_+$ will only weakly drive $\Phi_-$, 
 and any finite $\Phi_-$ will only weakly drive $\Phi_+$.
Further, 
 the two components evolve very differently, 
 one ``forwards'' in space at $\omega \sim mc^2/\hbar$
 and the other ``backwards'' at $\omega \sim -mc^2/\hbar$.
Thus any finite cross-coupling that does occur
 will be very poorly phase matched, 
 and will almost certainly average out to zero\footnote{See 
  appendix B of \cite{Kinsler-2010pra-fchhg}, 
  and also e.g. \cite{Kinsler-2007josab}.}.
This smallness criteria, 
 viz. 
~
\begin{align}
 \left|
  \left[
    \Xi-\frac{\hbar^2 \grad^2}{2m^2c^2}
  \right]
  \Phi_\mp
 \right|
&\ll
 \left|
  \left[
    1 \pm \frac{V}{mc^2}
  \right]
  \Phi_\pm
 \right|
,
\end{align}
 is therefore the minimum criteria which must hold
 for the Schr\"odinger equation to be valid; 
 although we should also be sure that
 any periodicities in $\Xi(\Vec{r},t)$ or $V(\Vec{r})$
 do not phase match the cross-coupling terms
 and allow them to accumulate to a significant level.

Assuming for now that this is true,
 as is indeed likely for non-relativistic low-momentum situations, 
 we get 
~
\begin{align}
  \imath
  \hbar
  \partial_t
  \Phi_\pm(\Vec{r})
&=
 \pm
  mc^2
  \Phi_\pm(\Vec{r})
 +
  V 
  \Phi_\pm(\Vec{r})
\nonumber
\\
&\qquad
 \pm
    m c^2 \Xi
  \Phi_\pm(\Vec{r})
 \mp
    \frac{\hbar^2 \grad^2}
         {2m}
  \Phi_\pm(\Vec{r})
.
\label{eqn-tprop-schrodinger-with-mass}
\end{align}

Next we can choose to --
 but are not compelled to -- 
 factor out the fixed rest-mass part, 
 which gives rise to fast oscillations induced by the energy
 of the particle's rest mass $m$.
This is done by introducing 
~
\begin{align}
  \Phi_\pm(\Vec{r})
&= 
  \psi_\pm(\Vec{r})
  e^{\pm \imath mc^2 t/\hbar}
,
\end{align}
 so that 
~
\begin{align}
  \imath
  \hbar
  \partial_t
  \psi_\pm(\Vec{r})
&=
 +
  V 
  \psi_\pm(\Vec{r})
 \pm
    m c^2 \Xi
  \psi_\pm(\Vec{r})
 \mp
    \frac{\hbar^2 \grad^2}
         {2m}
  \psi_\pm(\Vec{r})
.
\end{align}

Then
 we can choose our preferred direction --
 forwards in time -- 
 as indicated by a choice of upper signs, 
 so that 
~
\begin{align}
  \imath
  \hbar
  \partial_t
  \psi_{+}(\Vec{r})
&=
 +
  \left[
    V(\Vec{r}) + m c^2 \Xi(\Vec{r},t)
  \right]
  \psi_{+}(\Vec{r})
 -
    \frac{\hbar^2 \grad^2}
         {2m}
  \psi_{+}(\Vec{r})
,
\label{eqn-tprop-finalSchrodinger}
\end{align}
 which is the usual expression for the Schr\"odinger equation; 
 and we see that the effect of both Salpeter and gravitational potentials
 ends up essentially the same in this limit.
Since this derivation of the Schr\"odinger equation
 is for cases where the rest mass $m$ is dominant, 
 we might denote it the Schr\"odinger ``$m$-equation''.

If we were to consider propagating the wavefunction $\psi_{+}$
 forward in time, 
 we might divide both sides by $\imath \hbar$ to get
~
\begin{align}
  \partial_t
  \psi_{+}(\Vec{r})
&=
 -
  \frac{\imath}{\hbar}
  \left[
    V(\Vec{r}) + m c^2 \Xi(\Vec{r},t)
  \right]
  \psi_{+}(\Vec{r})
 +
    \frac{\imath \hbar \grad^2}
         {2m}
  \psi_{+}(\Vec{r})
.
\label{eqn-tprop-finalSchrodinger2}
\end{align}

It is worth noting that the last term
 in eqn. \eqref{eqn-tprop-finalSchrodinger}
 (or indeed eqn. \eqref{eqn-tprop-finalSchrodinger2})
 is a diffusion term, 
 and causes wavefunctions to spread outwards.
While this is the usually expected behaviour, 
 it is worth noting that being a diffusion does generate a causal problem --
 if starting from a strictly bounded wavefunction, 
 the diffusion term immediately generates some
 non-zero wavefunction values at arbitrarily large distances.
Thus parts of the wavefunction have propagated faster than lightspeed!
Of course, 
 this simply an artifact introduced by our mass-dominated 
 non-relativistic approximation; 
 it is not a feature of the initial Klein-Gordon wave equation, 
 which remains properly causal \cite{Kinsler-2011ejp}.
Having made such an approximation, 
 we should certainly not expect it to give useful
 (or even sensible)
 results for any effects propagating at or near lightspeed.
The artifacts are outside the scope allowed by the approximations used, 
 and however annoying, 
 they do not represent inherent physical failings.
If those artifacts are problematic in a particular case, 
 then the conclusion should be that 
 the Schr\"odinger equation is too approximate to use.

Lastly, 
 we can extract dispersion relations rather directly
 from eqns. \eqref{eqn-tprop-schrodinger-with-mass}
 or \eqref{eqn-tprop-finalSchrodinger2},
 by returing to the wavevector $\Vec{k}$ domain
 and using $\Vec{p} = \hbar \Vec{k}$.
It is
~
\begin{align}
  E - mc^2
&=
  V(\Vec{k}) + m c^2 \Xi(\Vec{k},t) + p^2/2m
,
\end{align}
 as expected in the mass-dominated limit considered.

%
% =======================================================================
\section{Momentum dominant: the $p$-equation}\label{S-tkinsler}

\begin{figure}
 \includegraphics[width=0.80\columnwidth,angle=0]{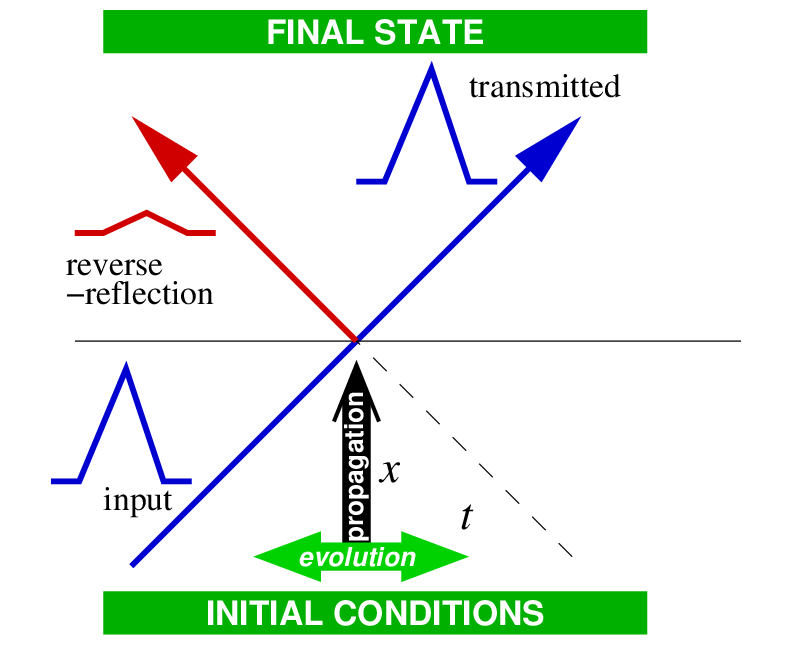}
\caption{
For spatial propagation, 
 initial conditions cover all times
 at an initial location $x_i$;
 the final state at $x_f$ again covers all times.
The effect of a reflective interface
 is also indicated, 
 since it makes the important distinction between 
 \emph{propagation} and \emph{evolution} clearer.
Note  that in this case, 
 any ``reflections'' generated are not
 like those we would normally expect, 
 although they are closely related.
This results from our insistence that all wave components 
 must travel (propagate) forward in space.
}
\label{F-diagram-zpropagator}
\end{figure}

In contrast to the intent of the Schr\"odinger derivation, 
 here we focus on momentum-dominated systems, 
 which naturally propagate with a strong spatial orientation.
%As a corollary to the momentum ($p$) dominance, 
% we will also find that the wavevector ($k$) is important.
This means we must aim to reduce the second-order-in-space KG equations
 down to the first-order-in-space 
 ``$p$-equation''
 by manipulating and approximating the starting equations 
 treating momentum as the quantity of primary importance.
Such a treatment typically makes most sense
 with very light or massless particles, 
 and indeed a spatially propagated description
 is very widely used in optics 
 (see e.g. \cite{Kinsler-2010pra-fchhg} and references therein).
This factorization 
 assumes a propagation forward in space, 
 whilst decomposing the system behaviour (waves) into components 
 that evolve either forward or backward in time, 
 as shown in fig. \ref{F-diagram-zpropagator}.
The consideration of spatial propagation 
 means that the result I present in this section is somewhat related 
 to the 
``spacelike counterpart of the Schr\"odinger equation'' 
 as previously derived by 
 by Holodecki \cite{Horodecki-1988inc}\footnote{Thanks
   to S.A.R. Horsley for the reference.}.
However, 
 the derivation presented here is \emph{necessarily} 
 aimed at the limit where
 the effect of the particle mass is a only small correction, 
 in contrast,
 Holodecki's result is limited to the non-relativistic regime.

To analyse spatial propagation, 
 we need a useful reference parameter to characterise it,
 and it should preferably be one that remains constant.
In this case, 
 a spatial frequency domain analysis is called for:
 we might therefore use either linear momentum $\Vec{p}$ 
 or a wavevector $\Vec{k}$.
This means that the parts of the physics we wish to ascribe 
 to the role of ``reference propagation'' must be independent
 of the primary propagation direction.

For the procedure to work, 
 we need to assume a direction along which the waves
 will primarily propagate.
Without loss of generality, 
 we will assume this to be the $z$-axis, 
 with the $x$ and $y$-axes to account for any transverse properties.
Thus we will focus on the $p_z$ momentum component, 
 and relegate $p_x$ and $p_y$ to the status of corrections.
After defining $\bar{E}^2 = E^2 - m^2c^4 = \hbar^2 \omega^2 - m^2c^4$, 
 and $p_T^2=p_x^2+p_y^2$, 
 we work in a spatial frequency (wavevector) space $\Vec{k}$.
Remembering that $\Vec{p} = \hbar \Vec{k}$, 
 %and for convenience writing any convolution $A \star B$
 %as $\hat{A} B$,
 we proceed in the following way
~
\begin{align}
&  \left[
    E - \hat{V}(\Vec{k})
  \right]^2 
  \Phi(\Vec{k},\omega)
\nonumber
\\
&\qquad =
  \left[
    m^2 c^4 
   +
    2 m^2 c^4 \hat{\Xi}(\Vec{k},\omega)
   +
    \hbar^2 c^2 k^2
  \right]
  \Phi(\Vec{k},\omega)
\\
&  \left[
    \hbar^2 c^2 k_z^2 
   -
    \left( E^2 - m^2 c^4 \right)
  \right]
  \Phi
\nonumber
\\
&\qquad =
  \left[
   -
    \hbar^2 c^2 k_T^2
   +
    \hat{V}^2
   -
    \left( E\hat{V} + \hat{V}E \right)
   -
    2m^2c^4\hat{\Xi}
  \right]
  \Phi \qquad
\end{align}
~
~
\begin{align}
  \left[
    \hbar^2 c^2 k_z^2 
   -
    \bar{E}^2
  \right]
  \Phi
&=
  \hat{W}
  \Phi
\\
  \left( \hbar c k_z - \bar{E} \right)
  \left( \hbar c k_z + \bar{E} \right)
  \Phi
&=
  \hat{W}
  \Phi
\\
  \Phi
&=
  \frac{1}
       {\left( \hbar c k_z - \bar{E} \right) \left( \hbar c k_z + \bar{E} \right)}
  \hat{W}
  \Phi
\\
  \Phi
&= 
  \left[
    \frac{1/2\bar{E}}
         {\hbar c k_z - \bar{E}}
   -
    \frac{1/2\bar{E}}
         {\hbar c k_z + \bar{E}}
  \right]
  \hat{W}
  \Phi
\label{eqn-zprop-mainderivH},
\end{align}
 where for convenience I have defined
~
\begin{align}
  \hat{W}(\Vec{k},\omega)
&=
  - c^2 p_T^2 + \hat{V}^2 - E\hat{V} - \hat{V}E - 2m^2c^4\hat{\Xi}
,
\end{align}
 retaining the ordering of $E$ and $V$
 as is needed when $E$ is (re)turned to operator form
 (i.e. as a time derivative).

We can see from the term in square brackets % $[]$
 on the RHS of eqn. \eqref{eqn-zprop-mainderivH}
 that $\Phi$ evolves according to two complementary parts
 of differing sign.
The term proportional to $(\hbar c k_z - \bar{E})^{-1}$
 generates a forward-like evolution, 
 and that proportional to $(\hbar c k_z + \bar{E})^{-1}$
 generates a backward-like evolution
 \cite{Ferrando-ZCBM-2005pre}.
As a result we can likewise split the wavefunction
 into matching pieces, 
 with $\Phi \equiv \Phi^+ + \Phi^-$.
When we transform back into the spatial domain, 
 these will (must!) propagate forward in space $z$, 
 all the while holding information about the wavefunction
 as a function of $x,y,t$.
To avoid notational clutter, 
 we use this as an excuse to omit the spatial argument $z$, 
 and only the $x,y,t$ arguments of $\Phi^\pm$ will be given.

Further, 
 since the $\Phi^+$ forward evolving component is by definition
 propagating to larger $z$, 
 it therefore must (also) be understood to 
 have excitations that evolve forward
 in time ($t$).
In contrast, 
 the $\Phi^-$ backward evolving component
 (also propagating to larger $z$),
 will contain excitations that evolve 
 backward in time.
While the notion of treating waves that evolve backward in time
 would (or perhaps should) typically be viewed with suspicion, 
 it can nevertheless be defended as a useful approximation in 
 many circumstances --
 notably, 
 this picture allows a remarkably powerful way of treating dispersion
 \cite{Kinsler-2012arXiv-fbacou}.

Continuing the separation of $\Phi^+$ and $\Phi^-$, 
 we see that 
~
\begin{align}
  \Phi^+(x,y,\omega) + \Phi^-(x,y,\omega)
&= 
  \left[
    \frac{1/2\bar{E}}
         {\hbar c k_z - \bar{E}}
   -
    \frac{1/2\bar{E}}
         {\hbar c k_z + \bar{E}}
  \right]
  \hat{W}
  \Phi(x,y,\omega)
\\
  \Phi^\pm
=
  \pm
    \frac{1/2\bar{E}}
         {\hbar c k_z \mp \bar{E}}
  \hat{W}
 &
  \left[
    \Phi^+(x,y,\omega) + \Phi^-(x,y,\omega)
  \right]
,
\end{align}
 and this enables us to write 
~
\begin{align}
  \left(
    \hbar c k_z \pm \bar{E}
  \right)
  \Phi^\pm(x,y,\omega)
&=
  \pm
    \frac{\hat{W}}
         {2\bar{E}}
  \left[
    \Phi^+(x,y,\omega) + \Phi^-(x,y,\omega)
  \right]
\\
  \hbar c k_z
  \Phi^\pm(x,y,\omega)
&=
 \pm
  \bar{E}
  \Phi^\pm(x,y,\omega)
\nonumber
\\
&\qquad
  \pm
    \frac{\hat{W}}
         {2\bar{E}}
  \left[
    \Phi^+(x,y,\omega) + \Phi^-(x,y,\omega)
  \right]
.
\end{align}
By again using the correspondance principle to 
 convert back from an $\omega, \Vec{k}$ based description, 
 into a $t,\Vec{r}$ form, 
 and with $\grad_T=(\partial_x,\partial_y,0)$,
 we get a pair of coupled differential equations,
~
\begin{align}
 -
  \imath
  \hbar c
  \partial_z
  \Phi^\pm(x,y,t)
&=
 \mp
  \imath
  \hbar
  \partial_t
  \Phi^\pm(x,y,t)
\nonumber
\\
&\quad ~
 \pm
    \frac{\hbar^2 c^2 \nabla_T^2}
         {2\bar{E}}
  \left[
    \Phi^+(x,y,t) + \Phi^-(x,y,t)
  \right]
\nonumber
\\
&\quad ~~
 \pm
    \frac{1}
         {2\bar{E}}
  \left[
    V^2
   -
    V \imath \hbar \partial_t
   -
    \imath \hbar \partial_t V
   -
    2 m^2 c^4 \Xi
  \right]
\nonumber
\\
&\qquad\qquad\qquad \times
  \left[
    \Phi^+(x,y,t) + \Phi^-(x,y,t)
  \right]
\\
  \partial_z
  \Phi^\pm
&=
 \pm
  c^{-1} \partial_t
  \Phi^\pm
 \pm
    \frac{\imath \hbar c \nabla_T^2}
         {2\bar{E}}
  \left(
    \Phi^+ + \Phi^-
  \right)
\nonumber
\\
&\quad ~
 \pm
    \frac{\imath}
         {2\hbar c \bar{E}}
  \left[
    V^2
   -
    \imath \hbar V \partial_t
   -
    \imath \hbar \partial_t V
   -
    2 m^2 c^4 \Xi
  \right]
\nonumber
\\
&\qquad\qquad\qquad \times
  \left[
    \Phi^+ + \Phi^-
  \right]
.
\label{eqn-zprop-pkequation}
\end{align}
On combining the time derivative terms,
 this becomes
~
\begin{align}
  \partial_z
  \Phi^\pm
&=
 \pm
  \frac{1}{c} %c^{-1} 
  \left[
    1
   +
    \frac{V}{\bar{E}}
  \right]
  \partial_t
  \Phi^\pm
 \pm
    \frac{V}{c \bar{E}}
  \partial_t
  \Phi^\mp
 \pm
    \frac{\imath \hbar c \nabla_T^2}
         {2\bar{E}}
  \left(
    \Phi^+ + \Phi^-
  \right)
\nonumber
\\
&\qquad 
 \pm
    \frac{\imath}
         {2\hbar c \bar{E}}
  \left[
    V^2
   -
    \imath \hbar \left( \partial_t V \right)
   -
    2 m^2 c^4 \Xi
  \right]
  \left(
    \Phi^+ + \Phi^-
  \right)
.
\label{eqn-zprop-pkequation2}
\end{align}

Whichever of
 eqns. \eqref{eqn-zprop-pkequation} or \eqref{eqn-zprop-pkequation2}
 you might prefer, 
 either consists of a pair of first order wave equations coupled 
 only by the potentials $V$ and $\Xi$,
 as scaled by the mass-compensated energy component ($\bar{E}$).
Those couplings, 
 along with the wavefunction(s), 
 then tell us how $\Phi^\pm$ will change on propagating forward in space $z$.
Note that unlike in the Schr\"odinger ($m$-) equation case, 
 there is consequently no explicit mass dependent oscillation; 
% $\pm (\imath mc^2/\hbar)t$ 
 the effect of the mass appears solely as a correction 
 to the effect of the potential; 
 although the rest-mass oscillation remains a legitimate contribution
 to $\Phi^\pm$.

If this potential-based coupling is small, 
 as is perhaps likely for light or massless particles, 
 then any finite $\Phi^+$ will only weakly drive $\Phi^-$, 
 and any finite $\Phi^-$ will only weakly drive $\Phi^+$.
Further, 
 the two components evolve very differently, 
 one ``forwards'' in time at $p \sim E/c$ % $k \sim +\omega/ c$
 and the other ``backwards'' at $p \sim - E/c$. % $k \sim -\omega/ c$.
Thus any finite cross-coupling that does occur
 will be very poorly phase matched, 
 and will almost certainly average out to zero.
This smallness criteria, 
 viz. 
~
\begin{align}
&
 \left|
  \left\{
    \frac{V}{\bar{E}}
   +
    \frac{\imath \hbar c^2 \grad_T^2}{2\bar{E}}
   +
    \frac{
          \imath
          \left[
            V^2
           -
            \imath \hbar \left( \partial_t V \right)
           -
            2 m^2 c^4 \Xi
          \right]
          }
         {2\hbar \bar{E}}
  \right\}
  \Phi_\mp
 \right|
\nonumber
\\
&\qquad\qquad\qquad
\ll
 \left|
  \left[
    1 + \frac{V}{\bar{E}}
  \right]
  \Phi_\pm
 \right|
,
\end{align}
 is therefore the minimum criteria which must hold
 for this $p$-equation to be valid; 
 although we should also be sure that
 any periodicities in $\Xi(\Vec{r},t)$ or $V(\Vec{r})$ 
 do not phase match the cross-coupling terms
 and allow them to accumulate significantly.

Assuming for now that this is true,
 as is indeed it might be for energetic but low-mass objects, 
 we get 
~
\begin{align}
  \partial_z
  \Phi^\pm
&=
 \pm
  c^{-1} \partial_t
  \Phi^\pm
 \pm
    \frac{\imath \hbar c \nabla_T^2}
         {2\bar{E}}
  \Phi^+
\nonumber
\\
&\qquad
 \pm
    \frac{\imath}
         {2\hbar c \bar{E}}
  \left[
    V^2
   -
    \imath \hbar V \partial_t
   -
    \imath \hbar \partial_t V
   -
    2 m^2 c^4 \Xi
  \right]
  \Phi^\pm
.
\end{align}
Or, 
 with combined time derivatives,
~
\begin{align}
  \partial_z
  \Phi^\pm
&=
 \pm
  \frac{1}{c}
  \left[
    1
   +
    \frac{V}{\bar{E}}
  \right]
  \partial_t
  \Phi^\pm
 \pm
    \frac{\imath \hbar c \nabla_T^2}
         {2\bar{E}}
  \Phi^+
\nonumber
\\
&\qquad
 \pm
    \frac{\imath}
         {2\hbar c \bar{E}}
  \left[
    V^2
   -
    \imath \hbar \left(\partial_t V\right)
   -
    2 m^2 c^4 \Xi
  \right]
  \Phi^\pm
.
\label{eqn-zprop-pkequation3}
\end{align}
In this last form, 
 we see that the typical (or ``reference'') wavevector $K$
 for a wavefunction component evolving with frequency $\omega$
 is 
~
\begin{align}
  K(\omega) 
&=
  \frac{\omega}{c}
  \left[ 
    1 + V(\omega/c) ~/~ \bar{E}
  \right] 
.
\end{align}

Again we have a diffusion-like term
 in our first order wave equation
 \eqref{eqn-zprop-pkequation3},
 here dependent on $\grad_T^2$.
Now, 
 however, 
 because we are propagating 
 a wavefunction known as a function of time
 forward in space,
 the extremes of the diffusion behaviour 
 (which is in this case actually a \emph{diffraction})
 correspond to very slow processes, 
 and therefore are not acausal artifacts\footnote{
  Note that the insistence on spatial propagation,
   however useful we may find it, 
   has already imposed some acausality 
   as a consequence of its nature.}.

%
% =======================================================================
\section{Summary}\label{S-summary}

I have shown how to derive the Schr\"odinger equation
 for a particle of mass $m$, 
 starting
 from the Klein-Gordon equation, 
 while taking into account the possible effects 
 of both static and/or dynamic potential landscapes
 influencing the evolution of the wavefunction.
This ``$m$-equation''
 is found using 
 an approximation which assumes that 
 the object's energy is dominated by its rest mass --
 i.e. that it is moving as non-relativistic speed.
The method is an adaption \cite{Kinsler-2012arXiv-fbacou} 
 of a factorization scheme
 recently applied in optics
 \cite{Kinsler-2010pra-fchhg,Kinsler-2012arxiv-fbrad,Kinsler-2007josab}, 
 but not originating from there.

Further, 
 I also derive an alternative to the Schr\"odinger equation
 in a different and complementary limit, 
 i.e. that of large momentum $p$.
This equation does not propagate the wave equation 
 forward in time, 
 as the usual Schr\"odinger equation does, 
 but forward in space.
This alternate ``$p$-equation''
 is presented here primarily as an exercise in technique, 
 and discussions of its possible utility 
 are left for later work.

\vspace*{70mm}

%
% =======================================================================
%\bibliography{/home/physics/_work/bibtex.bib}

\end{document}